\RequirePackage{fix-cm}
\RequirePackage{fixltx2e}
\documentclass[twoside,english,aip, jcp,floatfix]{revtex4-1}
\usepackage{mathpazo}

\usepackage[T1]{fontenc}
\usepackage[latin9]{inputenc}
\usepackage[a4paper]{geometry}
\usepackage{xcolor}
\usepackage{babel}
\usepackage{amsmath}
\usepackage{amssymb}
\usepackage{graphicx}
\usepackage[unicode=true,pdfusetitle,
 bookmarks=false,
 breaklinks=true,pdfborder={0 0 0},pdfborderstyle={},backref=false,colorlinks=true]
 {hyperref}
\hypersetup{
 citecolor = blue, linkcolor = blue,  urlcolor  = blue}

\makeatletter

\usepackage{orcidlink}

\makeatother

\begin{document}

\title{{\Large{}Stratification of uncertainties recalibrated by isotonic
regression and its impact on calibration error statistics }}

\author{Pascal PERNOT \orcidlink{0000-0001-8586-6222}}

\affiliation{Institut de Chimie Physique, UMR8000 CNRS,~\\
Université Paris-Saclay, 91405 Orsay, France}
\email{pascal.pernot@cnrs.fr}

\begin{abstract}
\noindent \emph{Post hoc} recalibration of prediction uncertainties
of machine learning regression problems by isotonic regression might
present a problem for bin-based calibration error statistics (e.g.
ENCE). Isotonic regression often produces stratified uncertainties,
i.e. subsets of uncertainties with identical numerical values. Partitioning
of the resulting data into equal-sized bins introduces an aleatoric
component to the estimation of bin-based calibration statistics. The
partitioning of stratified data into bins depends on the order of
the data, which is typically an uncontrolled property of calibration
test/validation sets. The tie-braking method of the ordering algorithm
used for binning might also introduce an aleatoric component. I show
on an example how this might significantly affect the calibration
diagnostics.
\end{abstract}
\maketitle

\section{Introduction}

\noindent \emph{Post hoc} recalibration of machine learned (ML) prediction
uncertainty for regression problems is currently an essential step
to achieve reliable uncertainty quantification (UQ)\citep{Tran2020}.
\emph{Post hoc} recalibration methods include, for instance, temperature
scaling\citep{Guo2017,Kuleshov2018,Levi2022}, isotonic regression\citep{Busk2022}
and conformal inference\citep{Angelopoulos2021,Hu2022}. 

Isotonic regression\citep{Barlow1972a} is used to scale prediction
uncertainties by an order-preserving non-linear function. According
to Busk \emph{et al.}\citep{Busk2022}, an isotonic regression model
$f_{\phi}(.)$ is designed to fit the empirical squared errors on
a held out calibration dataset. The recalibration function $f_{\phi}(.)$
takes as input the uncalibrated squared uncertainty $u_{i}^{2}$ and
outputs the scaled uncertainty $s_{i}^{2}u_{i}^{2}$ according to
\begin{equation}
s_{i}^{2}u_{i}^{2}=f_{\phi}(u_{i}^{2})
\end{equation}
Recalibration by isotonic regression has been shown to be efficient\citep{Busk2022,Busk2023_arXiv},
but it comes with a notable side effect. As the isotonic model $f_{\phi}(.)$
is piece-wise constant, the resulting calibrated uncertainties are
stratified.

I show in this note that such a stratification of uncertainty values
is problematic for the estimation of bin-based calibration statistics,
as the stratification makes the ordering of uncertainties \emph{ad
hoc} and ambiguous. For this, I use a recent test dataset ($M=13885)$
of predicted atomization energies and uncertainties from Busk \emph{et
al.}\citep{Busk2022} where the uncertainties have been recalibrated
by isotonic regression.

The next section (Sect.\,\ref{sec:Methods}) presents the calibration
metrics used to evaluate the impact of stratification. Application
to the dataset is presented in Sect.\,\ref{sec:Application}, leading
to the discussion of the results and conclusions in Sect.\,\ref{sec:Discussion-and-conclusion}.

\section{Methods\label{sec:Methods}}

\subsection{Calibration metrics\label{subsec:Validation-metrics}}

\noindent Let us consider a set of prediction errors and uncertainties
$\left\{ E_{i},u_{i}\right\} _{i=1}^{M}$. Average calibration can
be tested by comparing the RMV/MSE ratio or $\mathrm{Var}(Z)$ to
1 for the whole dataset\citep{Pernot2022b}, but in order to estimate
calibration errors, one needs metrics based on binned data. To do
so, the dataset is ordered by increasing uncertainty values and split
into $N$ equal-size bins $B_{1},\ldots,B_{N}$. Two calibration error
metrics are considered here, ENCE and ZVE:
\begin{itemize}
\item the Expected Normalized Calibration Error (ENCE)\citep{Levi2022}
\begin{equation}
\mathrm{ENCE}=\frac{1}{N}\sum_{i=1}^{N}\frac{|MV_{i}^{1/2}-MSE_{i}^{1/2}|}{MV_{i}^{1/2}}
\end{equation}
is a popular consistency metric that can be easily applied also to
adaptivity by a simple change of binning scheme.
\item the Z-Variance Error (ZVE)\citep{Pernot2023a_arXiv} is the mean deviation
from 1 of the binned $z$-scores variance
\begin{equation}
\mathrm{ZVE}=\exp\left(\frac{1}{N}\sum_{i=1}^{N}|\ln v_{i}|\right)
\end{equation}
accounting for the fact that the variance of \emph{z}-scores is a
dimensionless scale parameter
\begin{equation}
v_{i}=\mathrm{Var}\left(\left\{ Z_{j}=E_{j}/u_{j}\right\} _{j\in B_{i}}\right)\label{eq:LZV}
\end{equation}
\end{itemize}

\subsection{Calibration validation\label{subsec:Calibration-validation} }

\noindent Note that the ENCE and ZVE statistics are useful for comparisons
(the closer to their target the better) but they do not come with
a validation threshold and do not enable to ascertain or test perfect
calibration. This is due to the fact that they are based on a mean
absolute deviation (MAD) statistic which estimates a bias for biased
data and a dispersion for unbiased data.\citep{Pernot2018,Pernot2023a_arXiv} 

I have recently shown\citep{Pernot2023a_arXiv} that the search for
an optimal binning scheme to estimate these statistics is irrelevant.
If a dataset is strongly ill-calibrated, the ENCE or ZVE are practically
independent on the number of bins. The within-bin uncertainty is compensated
by the number of bins when averaging. However, for well or nearly
well-calibrated datasets, the MAD statistic involved in the ENCE and
ZVE definitions estimates the \emph{dispersion} of the binned values,
not their bias which is practically null. In consequence, the ENCE
and ZVE values are sensitive to the within-bin uncertainty, and both
are expected to be linear functions of $N^{1/2}$. In consequence,
there is no best value for $N$, and I proposed to use as a bin-independent
calibration metric the intercept of a linear interpolation of statistics
estimated for a series bin numbers. These intercepts are called ENCE$_{0}$
and ZVE$_{0}$. The 95\,\% confidence interval on the intercept is
used to validate calibration: the interval should contain 0 for ENCE$_{0}$
and 1 for ZVE$_{0}$. 

Depending on the datasets, the ENCE and ZVE values for small numbers
of bins (typically $N^{1/2}<4-6$) present notable deviations from
the expected linearity and should not be included in the linear fit.
At the other extremity of the bin-size range, it is better to keep
only bins with at least 30 points.\citep{Pernot2022a} 

\section{Application\label{sec:Application}}

\subsection{The dataset}

\noindent I use here the QM9 validation dataset presented in Busk
\emph{et al.}\citep{Busk2022}. It contains M =13\,885 predicted
atomization energies ($V$) and uncertainties ($u_{V}$) and reference
values ($R$). These data are transformed to $E=R-V$ and\textcolor{violet}{{}
}$u=u_{V}$. The uncertainties result from a \emph{post hoc} recalibration
by isotonic regression. 

\subsection{Stratification and binning}

\begin{figure}[t]
\noindent \begin{centering}
\includegraphics[width=0.99\textwidth]{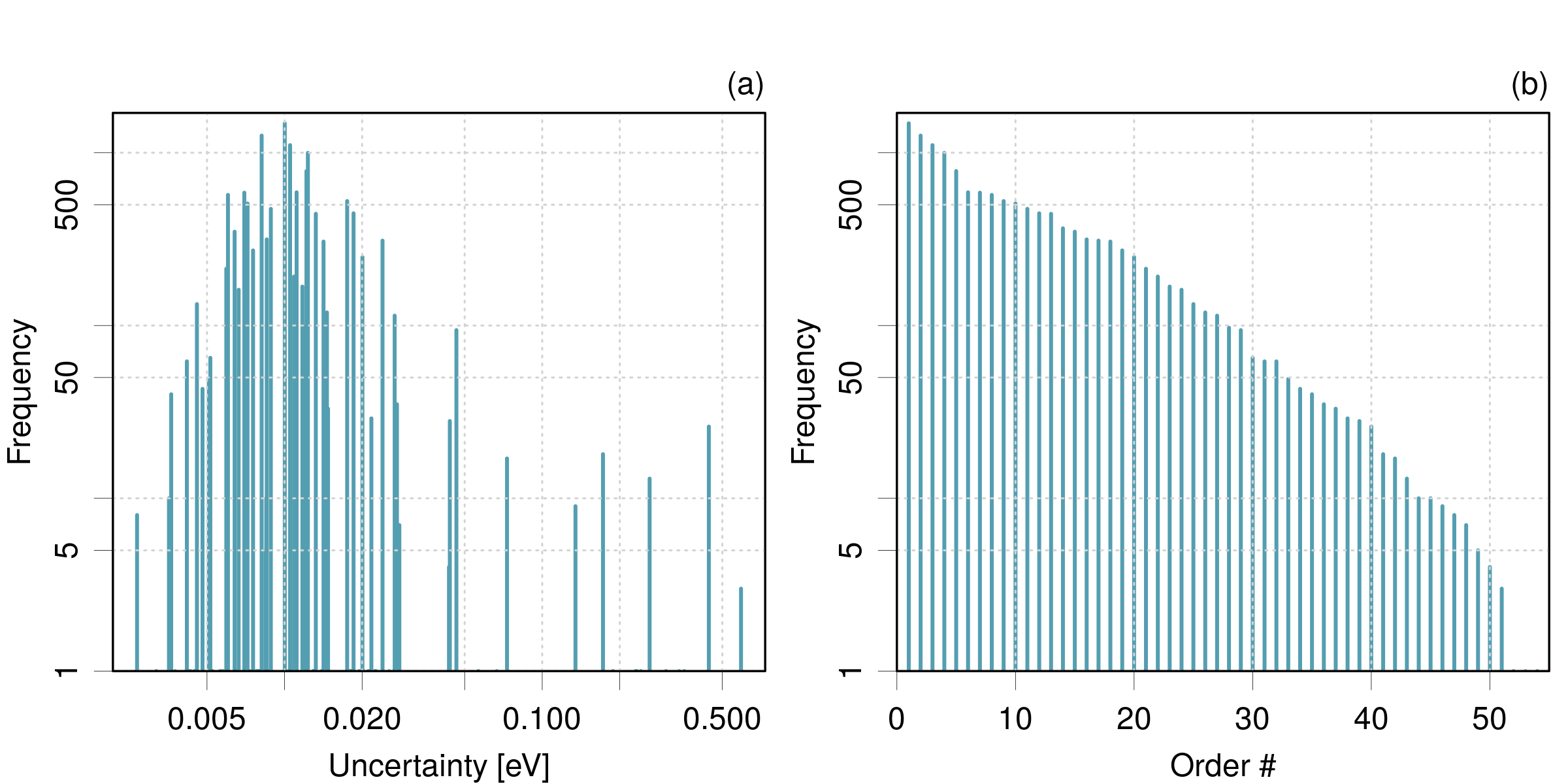}
\par\end{centering}
\caption{\label{fig:Stratification-statistics-of}Stratification statistics
of the uncertainties of QM9 validation dataset after recalibration
by isotonic regression.}
\end{figure}

\noindent The dataset contains only 138 unique uncertainty values,
87 of them being assigned to a single point. In consequence, most
uncertainty values are stratified, and 13\,798 points are distributed
among 51 uncertainty values. This distribution is shown in Fig.\,\ref{fig:Stratification-statistics-of}.
The stratification is mostly notable for uncertainty values between
0.005 and 0.02 eV {[}Fig.\,\ref{fig:Stratification-statistics-of}(a){]},
where the largest stratum contains about 1500 points. Ordering of
the strata by decreasing size {[}Fig.\,\ref{fig:Stratification-statistics-of}(b){]}
shows that there are ten strata with more than 500 points. 

For binning these data into equal-size bins, the data have first to
be ordered. Considering the number of ties, the resulting order might
depend firstly on how the ordering algorithms resolves ties and secondly
on the initial order of the data in the dataset. Once the data are
ordered, they are split into bins, which, depending on the number
of bins, might result into the fractioning of a large number of strata.
The problem does not lie into the splitting of the uncertainties themselves,
but into the ordering and splitting of the associated errors, which
can have a strong impact on the binned calibration statistics. 

Note that alternative binning schemes such as equal-width bins could
be considered, but they present other difficulties for the estimation
of error statistics (e.g. bins of very different sizes). 

\subsection{Data ordering and calibration statistics}

\begin{figure}[t]
\noindent \begin{centering}
\includegraphics[width=0.49\textwidth]{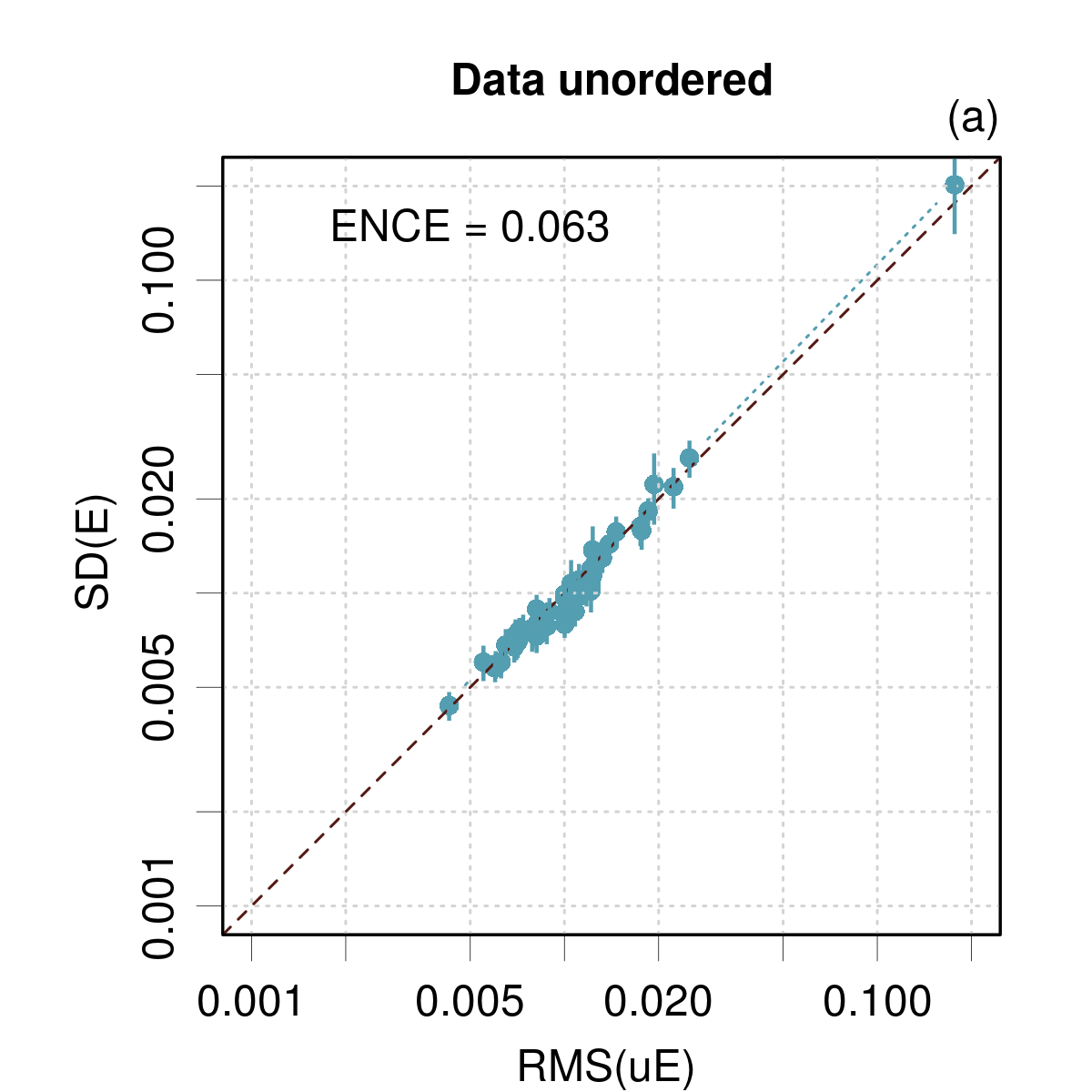}\includegraphics[width=0.49\textwidth]{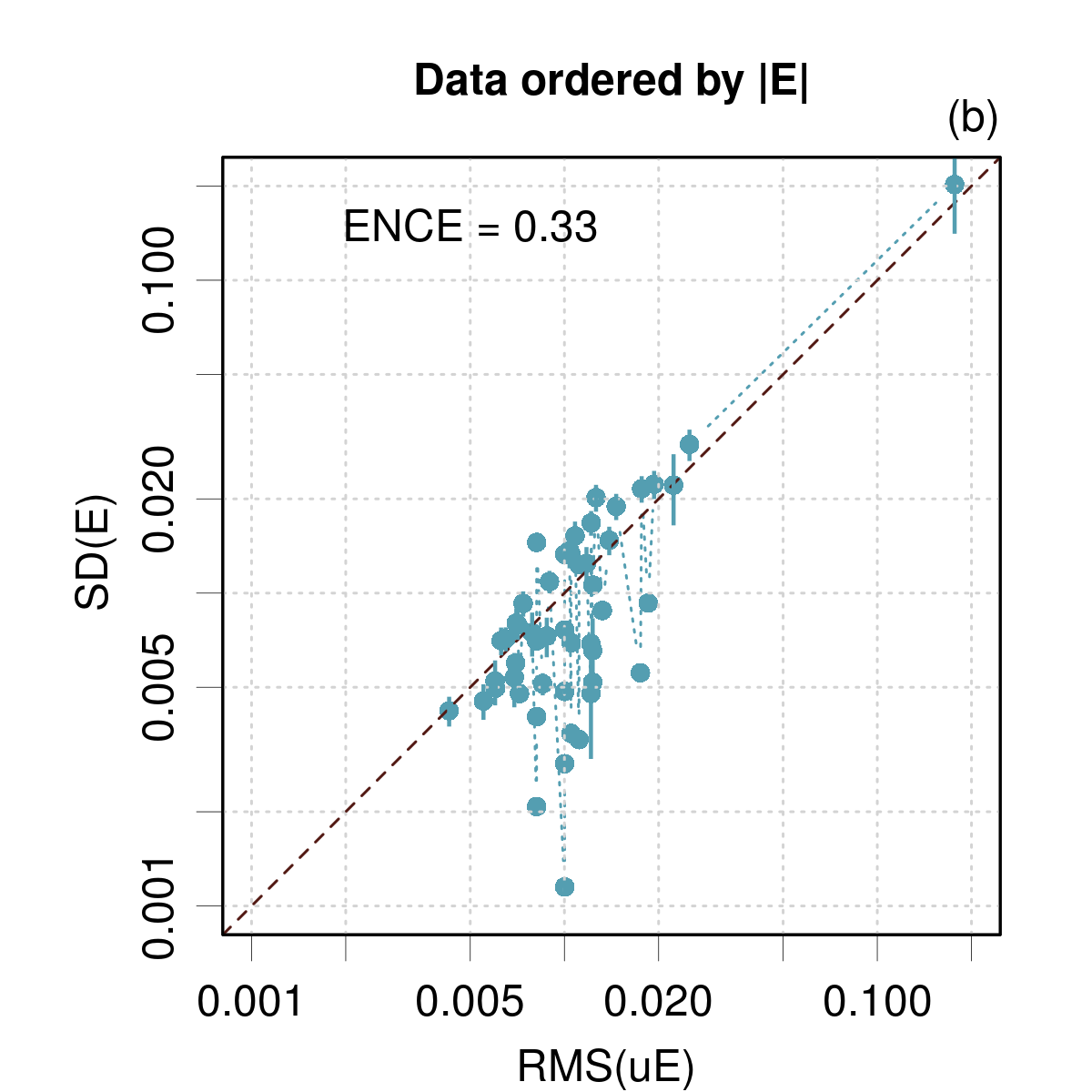}
\par\end{centering}
\caption{\label{fig:Stratification-statistics-of-2}Impact of data ordering
on reliability diagrams and ENCE values: (a) original data order;
(b) data reordered by increasing absolute errors. A large number of
bins (50) has been chosen to make the problem stand out.}
\end{figure}

\noindent As a worst case scenario, one can imagine that the data
associated to a given uncertainty value are ordered according to the
absolute errors. The gradient of absolute errors within a stratum
would result after splitting into bins with strong under- and over-estimated
uncertainties, and therefore very bad local calibration statistics
{[}Fig.\,\ref{fig:Stratification-statistics-of-2}(b){]}. By comparison,
the original order of the dataset (which is assumed to be random and
uncorrelated with $|E|$) results in a more reasonable reliability
diagram {[}Fig.\,\ref{fig:Stratification-statistics-of-2}(a){]}.
One can see also that the ENCE is strongly affected by the ordering,
from 0.063 for unordered data to 0.33 for ordered ones and $N=50$
bins. I chose here a large number of bins in order to accentuate the
effect (for small numbers of bins, the ordering is averaged out within
larger bins, but the error statistic is less accurate). For 15 bins,
which seems to be customary in the community, the ENCE still increases
from 0.05 to 0.13 upon ordering the data by $|E|$. 

A more plausible scenario is a random perturbation of the data order.
In fact this results into rather small fluctuations of the ENCE: 0.064(4)
($N=50$) where the mean and standard deviation are estimated from
a Monte Carlo sample of size 250. For the ZVE statistic in the same
conditions, one gets 1.14(1).

\begin{figure}[t]
\noindent \begin{centering}
\includegraphics[width=0.99\textwidth]{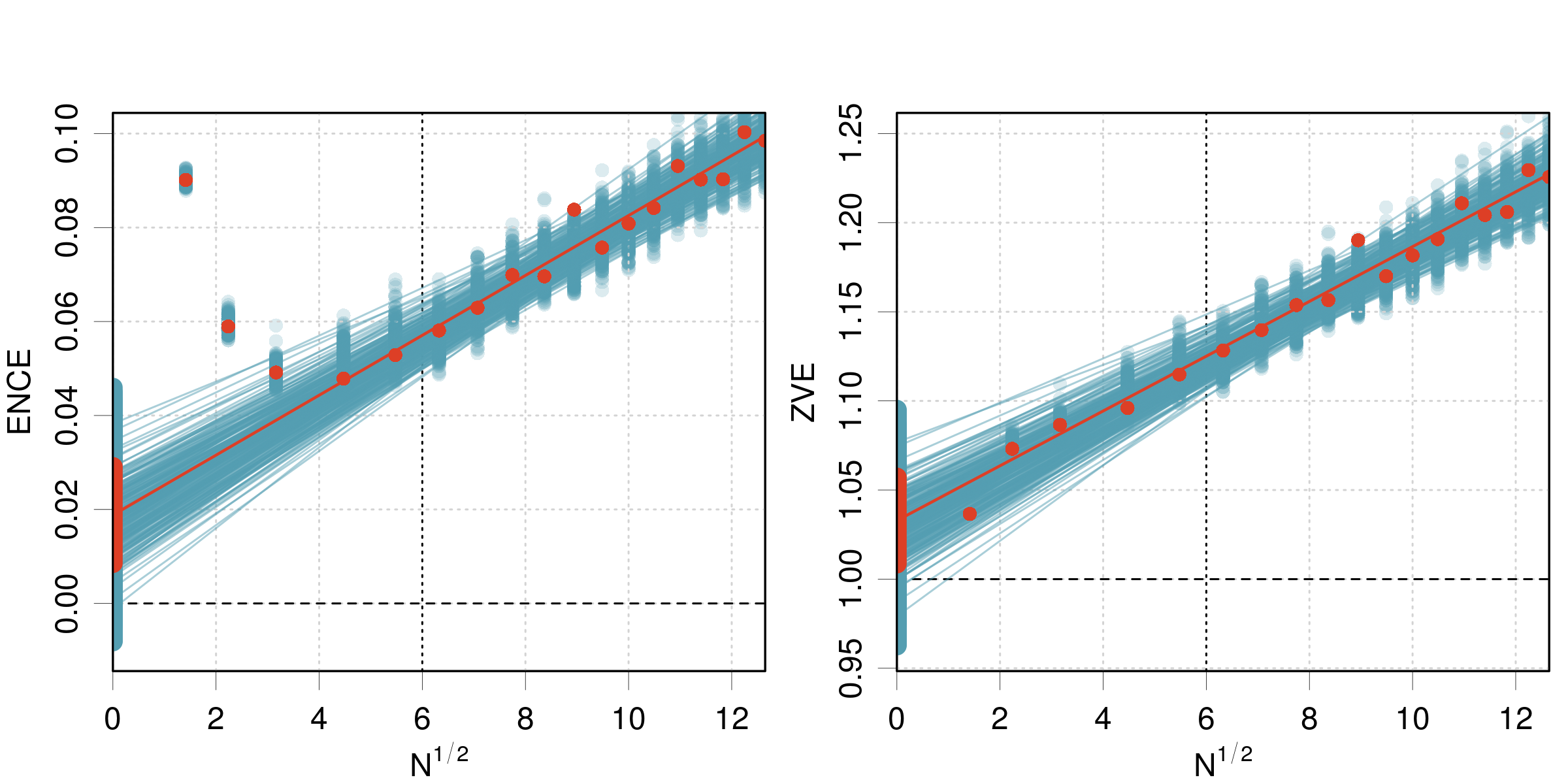}
\par\end{centering}
\caption{\label{fig:Stratification-statistics-of-1}Impact of the data ordering
on the ENCE$_{0}$ and ZVE$_{0}$ validation statistics.}
\end{figure}

As the values of ENCE and ZVE statistics depend on the number of bins,
bin-independent ENCE$_{0}$ and ZVE$_{0}$ calibration statistics
have also been used to evaluate the impact of data ordering. The results
can be seen on Fig.\,\ref{fig:Stratification-statistics-of-1}. The
ENCE and ZVE values for a series of bin numbers and the regression
lines using data with $N^{1/2}>6$ have been plotted for 250 random
orderings of the data (blue points and lines). The 95\,\% confidence
intervals for the intercept parameters have been plotted as thick
vertical segments. For comparison, the regression line and confidence
interval on ENCE$_{0}$ and ZVE$_{0}$ for the \emph{unperturbed}
dataset have been also reported (red points and lines). Although the
perfect calibration could be rejected by both statistics for the unperturbed
dataset (the 95\,\% confidence intervals do not contain the target
values), there is a non-negligible fraction of the randomly reordered
datasets for which validation is positive (about 8\,\% for ENCE$_{0}$
and 34\,\% for ZVE$_{0}$). 

\section{Discussion and conclusion\label{sec:Discussion-and-conclusion}}

\noindent We have seen that \emph{post hoc} calibration of prediction
uncertainties by isotonic regression poses a challenge for the estimation
of calibration metrics. This is due to the stratification of the calibrated
uncertainties which causes an ambiguity on the ordering of the uncertainties.
This ambiguity affects the binning of data for the estimation of calibration
statistics. Reordering the dataset by increasing absolute errors might
be catastrophic, whereas random reordering seems to have a more gentle
effect on calibration error statistics. However, I have shown that
calibration validation tests can be randomly affected by the choice
of ordering, leading to ambiguous conclusions. 

One has therefore to keep in mind that validation of the calibration
of a dataset with stratified uncertainties might be dependent on the
order of the data in the dataset. I am \emph{not} suggesting that
users permute their data until calibration is validated. Instead,
it would be nice if they could provide an estimation of the impact
of such permutations on their calibration error statistics. 

The use of \emph{centered} isotonic regression\citep{Oron2017} which
keeps the desirable properties of isotonic regression without the
stratification inconvenient would be a simple way to solve this problem. 

\section*{Acknowledgments}

\noindent I warmly thank J. Busk for providing me the data of the
BUS2022 case.

\section*{Author Declarations}

\subsection*{Conflict of Interest}

The authors have no conflicts to disclose.

\section*{Code and data availability\label{sec:Code-and-data}}

\noindent The code and data to reproduce the results of this article
are available at \url{https://github.com/ppernot/isotonic/releases/tag/v0.0}
and at Zenodo (\url{https://doi.org/10.5281/zenodo.8017641}). The
\texttt{R},\citep{RTeam2019} \href{https://github.com/ppernot/ErrViewLib}{ErrViewLib}
package implements the validation functions used in the present study,
under version \texttt{ErrViewLib-v1.7.0} (\url{https://github.com/ppernot/ErrViewLib/releases/tag/v1.7.0}),
also available at Zenodo (\url{https://doi.org/10.5281/zenodo.7729100}).\textcolor{orange}{{} }

\bibliographystyle{unsrturlPP}
\bibliography{NN}

\end{document}